\pdfoutput=1     %% Comment if you are not using pdflatex
\documentclass[12pt]{JINST}
\usepackage{amsmath,graphicx,epsfig,amssymb,dsfont,mathtools}
\usepackage{inputenc}
\usepackage{ulem} %% for strike-through
\usepackage{bigstrut}
\usepackage{makecell}
\usepackage{slashed}
\usepackage{multirow}
\usepackage{subfigure}
\usepackage{url}
\usepackage{diagbox}
\usepackage{etoolbox}
\usepackage{lineno}
\usepackage{diagbox}
\usepackage{rotating}
\usepackage{lscape}
\usepackage{array}
\maketitle
\linenumbers
\title{Energy reconstruction of hadronic showers at the CERN PS and SPS using the Semi-Digital Hadronic Calorimeter}

\author{\centering 
\LARGE\bf The CALICE Collaboration}

\author{\centering
D.\,Boumediene
\\ \it
Universit\'e Clermont Auvergne, Universit\'e Blaise Pascal, CNRS/IN2P3, LPC, 4 Av. Blaise Pascal, TSA/CS 60026,
F-63178 Aubi\`ere, France
}

\author{\centering
A.\,Pingault, 
M.\,Tytgat
\\ \it
Ghent University, Department of Physics and Astronomy,
Proeftuinstraat 86, B-9000 Gent, Belgium
}

\author{\centering Y.\,W.\,Baek, D-W.\,Kim\footnote{Now at Seoul National University Hospital}, S.\,C.\,Lee, B.\,G.\,Min, S.\,W\,Park \\ \it Gangneung-Wonju National University Gangneung 25457, South Korea }

\author{\centering 
Y.\,Deguchi,
K.\,Kawagoe,
Y.\,Miura,
R.\,Mori,
I.\,Sekiya,
T.\,Suehara,
T.\,Yoshioka
\\ \it
Department of Physics and Research Center for Advanced Particle Physics,
Kyushu University, 744 Motooka, Nishi-ku, Fukuoka 819-0395, Japan
}

\author{\centering 
L.\,Caponetto, 
C.\,Combaret, 
G.\,Garillot, 
G.\,Grenier, 
J-C.\,Ianigro, 
T. \,Kurca,  
I.\,Laktineh$^{a}$ , 
B.\,Liu$^{a}$ ,
B.\,Li,
N.\,Lumb, 
H.\,Mathez, 
L.\,Mirabito, 
A.\,Steen\footnote{Now at NTU}
\\ \it
Univ Lyon, Univ CLaude Bernard Lyon 1, 
CNRS/IN2P3, IP2I Lyon, F-69622 
Villeurbanne, France
}

\author{\centering 
E.\,Calvo Alamillo,
C.\, Carrillo,
M.C.\,Fouz,
H.\, Garcia Cabrera,
J.\,Marin,
J.\,Navarrete,
J.\,Puerta Pelayo,
A.\,Verdugo
\\ \it
CIEMAT, Centro de Investigaciones Energeticas, Medioambientales y Tecnologicas, Madrid, Spain 
}

\author{\centering 
F.\,Corriveau
\\ \it
Department of Physics, McGill University,
Ernest Rutherford Physics Bldg.,
3600 University Ave.,
Montr\'{e}al, Qu\'{e}bec
Canada H3A 2T8
}

\author{\centering 
L.\,Emberger,
C.\,Graf,
L.M.S\,de Silva,
F.\,Simon,
C.\,Winter
\\ \it
Max-Planck-Institut f\"ur Physik,
F\"ohringer Ring 6,
D-80805 Munich, Germany
}

\author{\centering 
J.\,Bonis, 
D.\,Breton, 
P.\,Cornebise, 
A.\,Gallas, 
J.\,Jeglot, 
A.\,Irles, 
J.\,Maalmi, 
R.\,P\"oschl, 
A.\,Thiebault, 
F.\,Richard, 
D.\,Zerwas 
\\ \it
Universit Paris-Saclay, CNRS/IN2P3, IJCLab, 91405 Orsay, France 
}

\author{\centering 
J.\,Cvach, 
M.\,Janata, 
M.\,Kovalcuk, 
J.\,Kvasnicka, 
I.\,Polak, 
J.\,Smolik, 
V.\,Vrba, 
J.\,Zalesak, 
J.\,Zuklin
\\ \it
Institute of Physics, The Czech Academy of Sciences,
Na Slovance 2, CZ-18221 Prague 8, Czech Republic
}

\author{\centering
Y.Y. \,Duan,
 S.  \,Li, 
 J.  \,Guo,
  J.F. \, Hu, 
  F.  \,Lagarde,
  B.\,Liu, 
  Q.P. \, Shen, 
  X.  \,Wang, 
  W.H. Wu, 
  H.J.  \,Yang, 
  Y.F.  \,Zhu
\\ \it 
Tsung-Dao Lee Institute, Institute of Nuclear and Particle Physics, School of Physics and Astronomy, Shanghai Jiao Tong University,
Key Laboratory for Particle Physics, Astrophysics and Cosmology (Ministry of Education),
Shanghai Key Laboratory for Particle Physics and Cosmology,
800 Dongchuan Road, Shanghai, 200240, P. R. China
}

\author{\centering
L.\,Emberger,
C.\,Graf,
L.M.S\,de Silva,
F.\,Simon,
C.\,Winter
\\ \it
Max-Planck-Institut f\"ur Physik,
F\"ohringer Ring 6,
D-80805 Munich, Germany
}

\author{
\it
$^{a}$ Corresponding authors\newline
E-mail: \email{laktineh@in2p3.fr, liubinglove7@gmail.com}

}

\abstract{The CALICE Semi-Digital Hadronic CALorimeter (SDHCAL)  is the first technological prototype in a family of high-granularity calorimeters developed by the CALICE Collaboration to equip the experiments of future lepton colliders.  The SDHCAL is a sampling calorimeter using stainless steel for absorber and Glass Resistive Plate Chambers (GRPC) as a sensitive medium. The GRPC are  read out by 1~cm $\times$ 1~cm pickup pads combined to a multi-threshold electronics.  The prototype was exposed to hadron beams in both the CERN PS and the SPS  beamlines in 2015 allowing the test of the SDHCAL in a large energy range from 3~GeV to 80~GeV.   After introducing the method used to select the hadrons of our data and reject the muon and electron contamination,  we present the energy reconstruction approach that we apply to the data collected from both beamlines and we discuss the response linearity and the energy resolution of the SDHCAL.   The results obtained in the two beamlines confirm the excellent SDHCAL performance observed  with the data collected with the same prototype in the SPS beamline in 2012. They also show  the stability of the SDHCAL in different beam conditions and different time periods.}

\author{Calice Collaboration}
\begin{document}

\label{sec:intro}
\section{Introduction}

The Semi-Digital Hadronic CALorimeter (SDHCAL)~\cite{SDHCAL} is the first of a series of technological high-granularity prototypes developed by the CALICE collaboration. The   SDHCAL  is a sampling calorimeter using stainless steel for absorber and Glass Resistive Plate Chambers (GRPC) for its sensitive medium. 
 The SDHCAL is designed to be as compact as possible with its mechanical structure being part of the absorber. The GRPC and the readout electronics are conceived to achieve minimal dead zones\~cite{SDHCAL}. This design renders the SDHCAL optimal for the application of the Particle Flow Algorithm (PFA) techniques~\cite{PFA1, PFA2, PFA3}. 

\begin{figure}[h]
\centerline{\includegraphics[width=0.70\columnwidth]{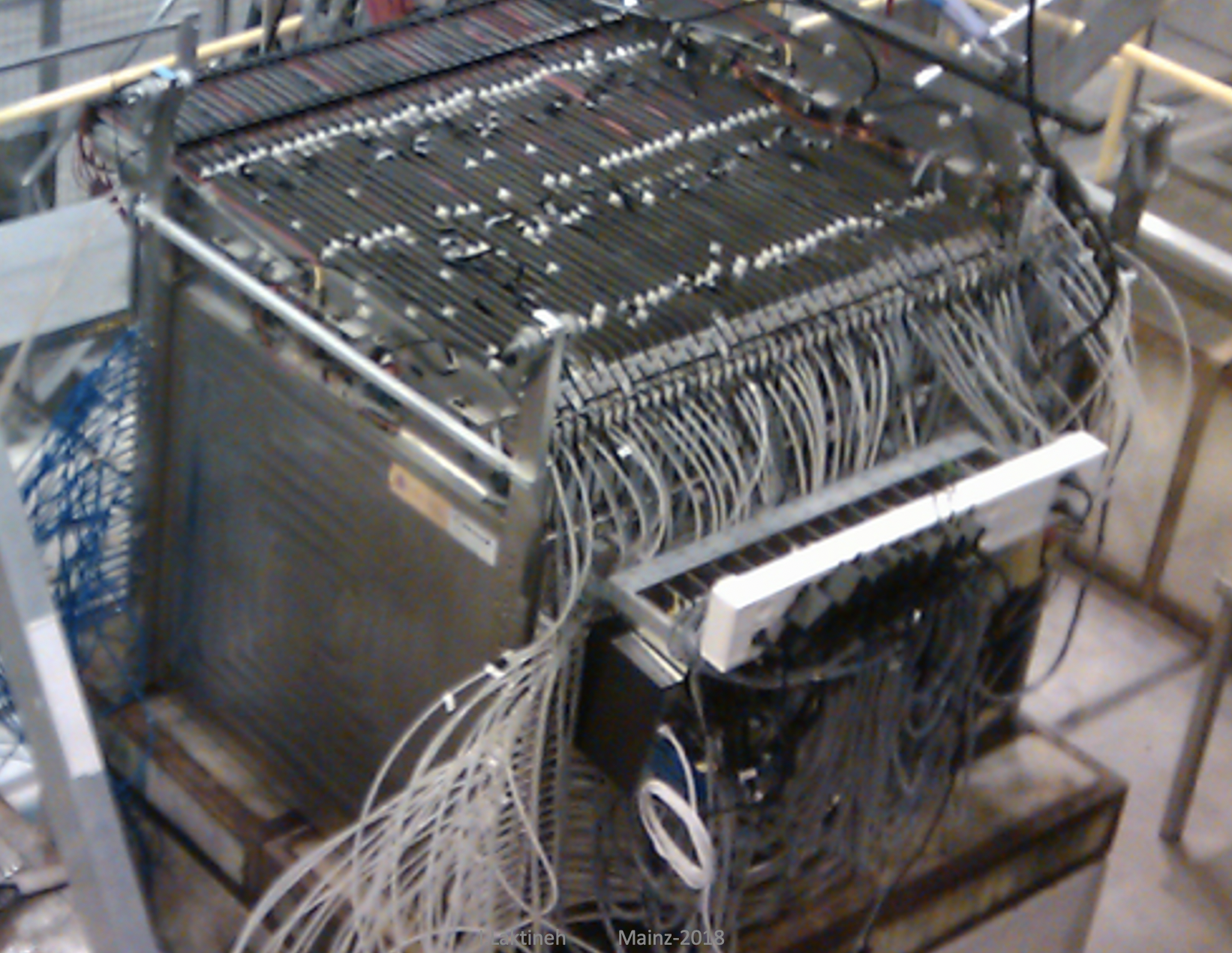}}
\caption{A picture of the SDHCAL prototype in beam test.}\label{picture}
%\end{wrapfigure}
\end{figure}

\begin{figure}[h]
\centerline{\includegraphics[width=0.70\columnwidth]{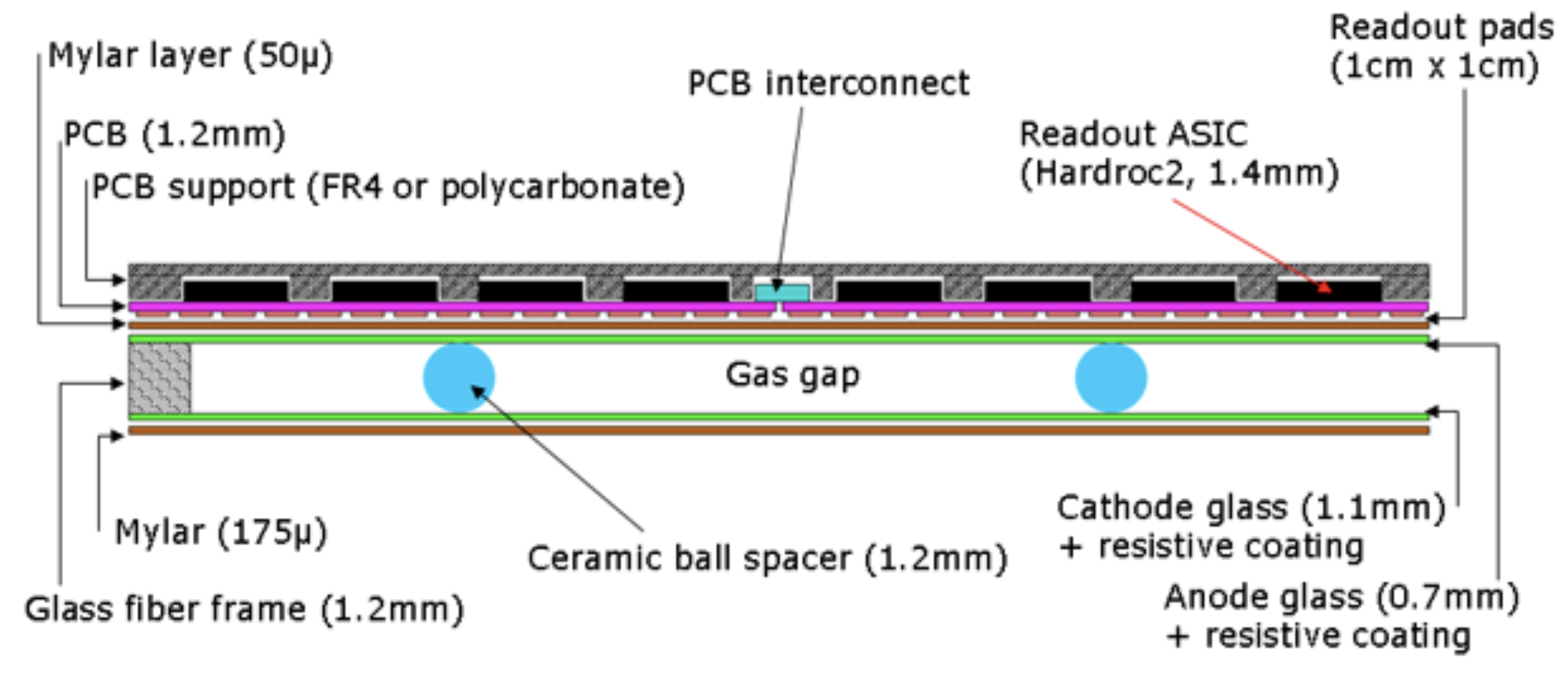}}
\caption{Cross-section of one of the prototype's 48 active  layers.}\label{scheme}
%\end{wrapfigure}
\end{figure}

 The SDHCAL (~\ref{picture}) comprises 48 active layers, each of them equipped with a 1~m~$\times$~1~m GRPC and  an Active Sensor Unit (ASU) of the same size hosting on one side (the one in contact with the GRPC)~pickup pads of 1~cm~$\times$~1~cm size each and 144 HARDROC2 ASICs~\cite{HARD} on the other side (Fig.~\ref{scheme}). The GRPC and the ASU are assembled within a cassette made of two stainless steel plates, 2.5~mm thick each. The 48 cassettes are inserted in a self-supporting mechanical structure made of 49 plates, 15~mm thick each, of the same material as the cassettes, bringing the total absorber thickness to 20~mm per layer.   The empty space between two consecutive plates is 13~mm to allow the insertion of one cassette of 11~mm thickness.  In total, the SDHCAL represents about 6 interaction lengths $\lambda_I$. 
   The HARDROC2 ASIC has 64 channels to read out 64 pickup pads. Each channel has three parallel digital circuits whose parameters can be configured to provide 2-bit encoded information per channel indicating if the charge seen by each pad has passed any of the three different thresholds associated to each digital circuit. This multi-threshold readout is used to improve on the energy reconstruction of hadronic showers at high energy ($> 30$~GeV) with respect to the simple binary readout mode as explained in Ref.~\cite{FirstResults}.
 
 The SDHCAL was exposed to different kinds of particles  at the CERN SPS beamline in 2012 and its performance was studied in the the energy range above 10~GeV~\cite{FirstResults} since lower particle energy values are difficult to obtain in the SPS beamline standard configurations. To study its performance in different beam conditions and at lower energies, the SDHCAL   was exposed to hadrons both in the SPS and PS beamlines in 2015. It was first exposed to negatively charged pion beams of 3, 4, 5, 6, 7, 8, 9, 10 and 11~GeV at the PS beamline and then to positively charged hadrons of 10, 20, 30, 40, 50, 60, 70 and 80~GeV at the SPS beamline. In both cases, about 10000 events were collected for each energy point.
 
 In this paper, section~\ref{simu} gives the details of the collected beam data as well as the samples of simulated events used for comparison. The pion selection and the muon and electron contamination rejection using the MultiVariate Analysis (MVA) technique known as  Boosted Decision Trees~(BDT)  to separate pion and election showers is  given in section~\ref{selection}. Energy reconstruction of the selected pion events is discussed in section~\ref{reco}. Finally, in  section~\ref{uncertain} we  present the uncertainties related to the energy reconstruction of the collected data.
 
 %\begin{wrapfigure}{r}{0.5\columnwidth}

%\begin{wrapfigure}{r}{0.5\columnwidth}

\section{Simulation}
\label{simu}
%In the year of 2015, The SDHCAL prototype was exposed to beams of hadrons, electrons and muons on the beamlines of PS in the CERN. In this period, the negatively charged pion beams of 3, 4, 5, 6, 7, 8, 9, 10 and 11~GeV were taken.  We also collect the positively hadrons  including the  10, 20, 30, 40, 50, 60, 70 and 80~GeV in the beamlines of SPS. In this paper, we only focus on the low energy hadrons beams taken at PS concerning the pion event selection, because of the hadron beams taken at SPS already studied in the ref.~\cite{RefCAN}.
%For such beam pion samples, the muon contamination exists in all samples. However, The electron contamination of the energy points including 3, 4 and 5~GeV need to be checked. For the the samples from 6 to 11~GeV, they don't have electron contamination thanks to the the use of electron eliminator.  
 The simulation model of SDHCAL, based on Geant4.9.6 toolkit package~\cite{GEANT4}, was developed including the interactions of different kinds of particles such as muons, electrons and pions in the SDHCAL prototype. The simulation takes into account the operation conditions of the GRPC to which an  effective high voltage of 7.2 kV  was applied. It uses the same values  of 0.114, 5 and 15 pC that are used  by the SDHCAL readout system for the first, the second and the third threshold respectively.

 Among the different Geant4 physics lists that were used to compare the simulation with the beam data collected in 2012,  FTF\_BIC was found to provide the best agreement~\cite{digitizer}. Therefore, we use this physics list in this work to simulate events with different kinds of particles having the same energies and impinging on the prototype in the same area as those to which the SDHCAL was exposed during the beam test campaigns at the  SPS and PS beamlines. The simulated events are then used
 to optimize the selection of  pions in data by rejecting muon and electron beam contamination. They are also used to estimate possible biases that may influence the  data energy reconstruction of hadronic showers in terms of linearity and resolution.

\section{Pion events selection}
\label{selection}
The pion samples of  both SPS and PS beams  are contaminated by two kinds of particles: electrons and muons.  The muon contamination exists in all samples from 3 to 80~GeV. This  includes two different types of muons: cosmic muons and beam muons. The latter are generated by pions decaying before arriving at the prototype. Concerning the electron contamination, it is negligible in pion samples from 6 to 11~GeV  in the PS  pion beam but still present in the energy range below 6~GeV at the level of a few percent of the beam content. % with  the highest contamination 3~GeV. 
%This contamination results from the way the  PS beam is produced~\ref{PS}. 
 The electron contamination is also present in the SPS pion beam, especially in the energy range between 10 and 50~GeV~\cite{FirstResults}.  The muon rejection is rather easy due to their track-like shape that distinguishes them clearly from the hadronic showers in the SDHCAL. The electron rejection is harder.  Electron showers, in particular at low energy,  are similar to the pion ones.  Although in both PS and SPS, an electron stopper made of  a few millimeters thick lead plate was used, this does not allow to completely eliminate the  electron contamination.  Inspired by Refs.~\cite{BDT,BDT1},   we propose to use  the BDT technique to reject the electron background of our pion samples in an improved way  with respect to the one used in a previous analysis applied to data collected by  SDHCAL in 2015 at the SPS beamline~\cite{RefCAN} in the energy range between 10 and 80 GeV.

\subsection{Electron contamination rejection using Boosted Decision Tree~(BDT)}
As mentioned in the previous section, for the 6 to 11~GeV pion runs, the electron contamination is negligible in the PS pion beam. However for the 3 to 5 GeV pion runs, the electron contamination is expected to be present due to the PS beamline structure. Therefore, it is necessary to check the electron contamination and to eliminate it.  
Thanks to the high granularity of the SDHCAL prototype, we can use the  BDT method to exploit the three dimensional shape  of both the electromagnetic and the hadronic showers to classify the electron and pion events in our prototype.  The BDT method is one of the most powerful and widely used in high energy physics for classification tasks. The TMVA package~\cite{MVAT} contains a standardised implementation of this technique. We adopt this package to build our BDT model to reject the electron contamination.

\subsubsection{ BDT input variables} Based on the difference in topology between electromagnetic and hadronic showers, we choose eight variables as inputs of the BDT model to help discriminate pion against electron events. Hereafter, a description of each of these variables is given:
 \begin{itemize}

\item {\bf  First layer of the shower (Begin)}: To define the layer in which the shower starts, we look for the first layer along the incoming particle direction which contains at least 4 fired pads. To eliminate fake shower starts due to accidental noise or a locally high multiplicity, the following 3 layers after the first layer are also required to have more than 4 fired pads (hits) for each of them.  If no layer fulfils this, a value of -10 is assigned to the variable. Since each layer  of the SDHCAL represents  about 1.2 radiation lengths ($X_0$),  electromagnetic showers start developing in the first layers. For pions, their interaction probability density is given by  $1-\exp{(-\frac{l}{\lambda_I})}$ where $l$ is the length of the pion trajectory in the calorimeter medium before interacting. Figure~\ref{begin} shows the distribution of the first layer of the shower in the SDHCAL prototype for pions and electrons as obtained from the simulation.

\begin{figure}[!h]
\begin{center}
\includegraphics[width=0.48\textwidth]{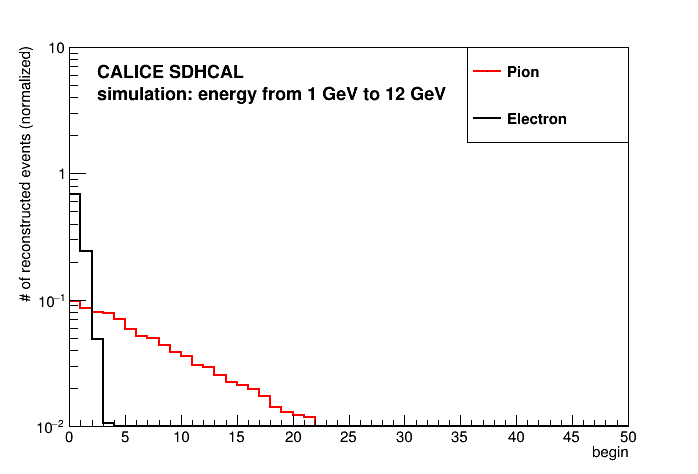}
\includegraphics[width=0.48\textwidth]{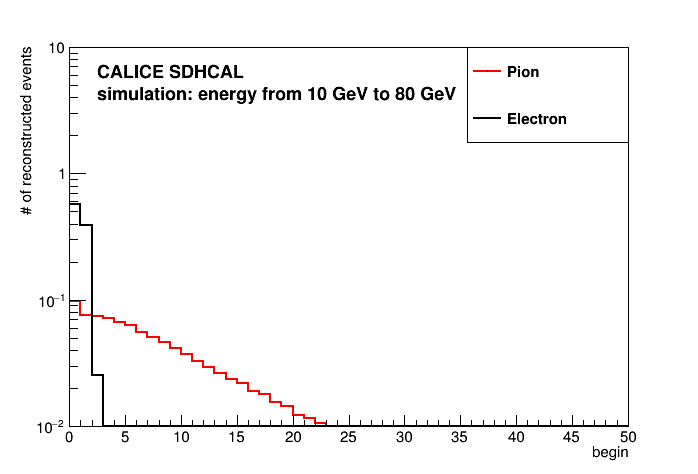}
\caption{\label{begin} Distribution of the first layer number of pion and electron showers in the PS energy range of 1-12~GeV (left) and the SPS energy range of 10-80~GeV (right) of pions and electrons as given by the simulation. The red line corresponds to  pions and the  black one to electrons.}
\end{center}
\end{figure}

\item {\bf Number of track segments in the shower (nTrack)}: Applying the Hough Transform~(HT) method to single out the tracks in each event as described in Ref.~\cite{HT}, we obtain the number of track segments in the pion, electron and muon events. A HT-based segment candidate is considered as a track segment if there are more than 6 aligned hits with not more than one layer separating two consecutive hits.  Electron showers feature almost no track segment while most of the hadronic showers have at least one. The distribution of nTrack can be seen in Fig.~\ref{nTrack}
\begin{figure}[!h]
\begin{center}
\includegraphics[width=0.48\textwidth]{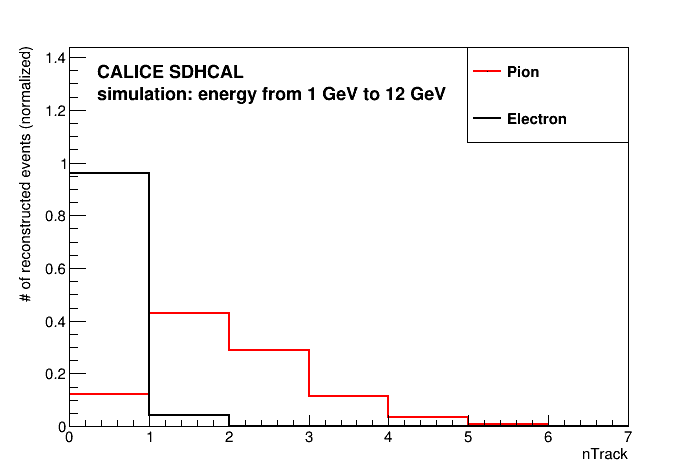}
\includegraphics[width=0.48\textwidth]{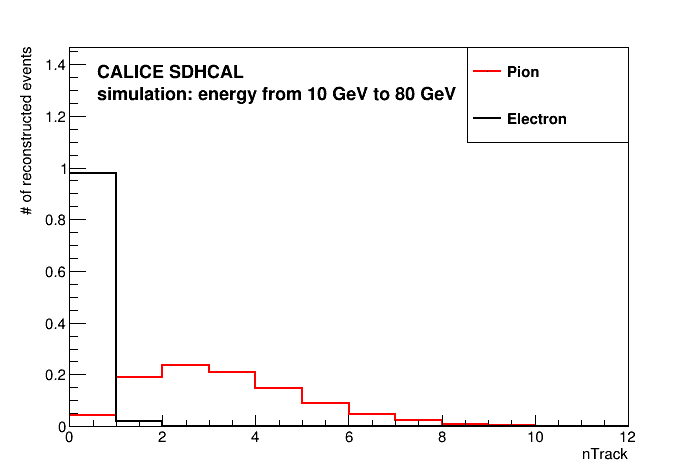}
\caption{\label{nTrack} Distribution of the number of track segments  in pion and electron showers in the PS energy range of 1-12~GeV (left) and the SPS energy of 10-80~GeV (right)  as given by the simulation. The red line corresponds to  pions and the  black one to electrons.}
\end{center}  
\end{figure}

\item {\bf Number of clusters of the shower (nCluster)}: All hits in a given layer are clustered using a nearest-neighboring algorithm described in Ref.~\cite{SDHCAL}. It consists in merging in each GRPC plate the hits sharing a common edge. This variable defines the number of clusters of the shower and its distribution is shown in Fig.~\ref{nCluster}.  Indeed, the compactness of the electromagnetic shower leads to a reduced number of clusters in an electron shower  with respect to that of a pion shower of the same energy.
\begin{figure}[!h]
\begin{center}
\includegraphics[width=0.48\textwidth]{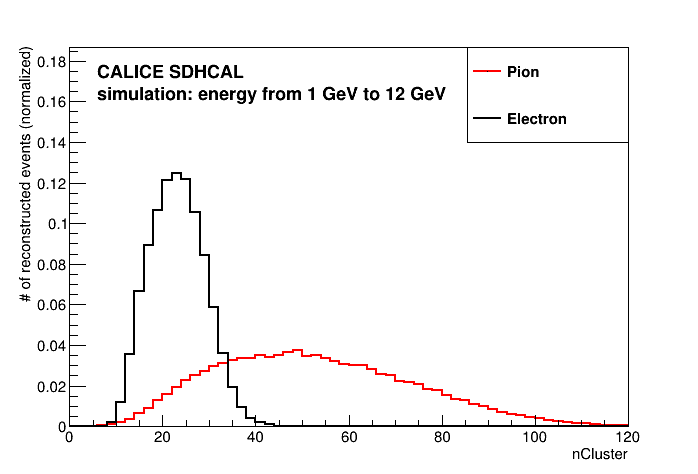}
\includegraphics[width=0.48\textwidth]{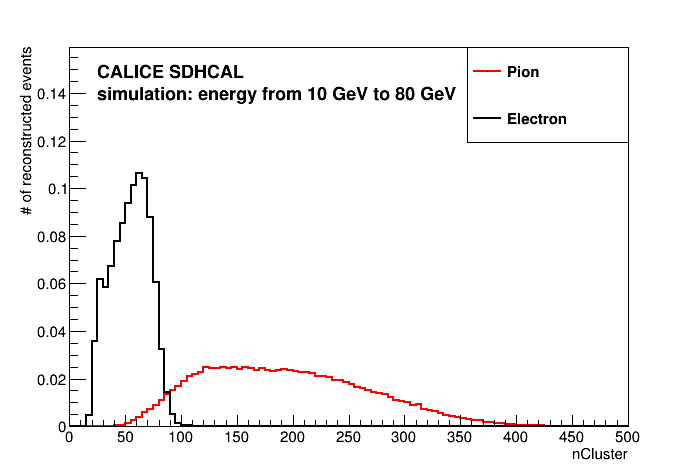}
\caption{\label{nCluster} Distribution of the number of clusters in pion and electron showers  in the PS energy range of 1-12~GeV (left) and  the SPS range of 10-80~GeV (right) as given by the simulation. The red line corresponds to pions and the black one to electrons.}
\end{center}
\end{figure}

\item {\bf Ratio of the number of shower layers over the total number of fired layers (nInteractingLayers/nLayers)}: This is the ratio between the number of the shower layers defined as those in which the Root Mean Square (RMS) of the hits' position in the $x$-$y$ plane exceeds 5~cm in both $x$ and $y$ directions and the total number of layers with at least one hit. %Although this variable is correlated with the shower start one, 
This variable allows, as can be seen in Fig.~\ref{Ratio2}, a good separation between pions and electrons at low energy. It also allows an easy discrimination of muons (including the radiative ones) against pions and electrons ones as will be shown later. 
\begin{figure}[!h]
\begin{center}
\includegraphics[width=0.48\textwidth]{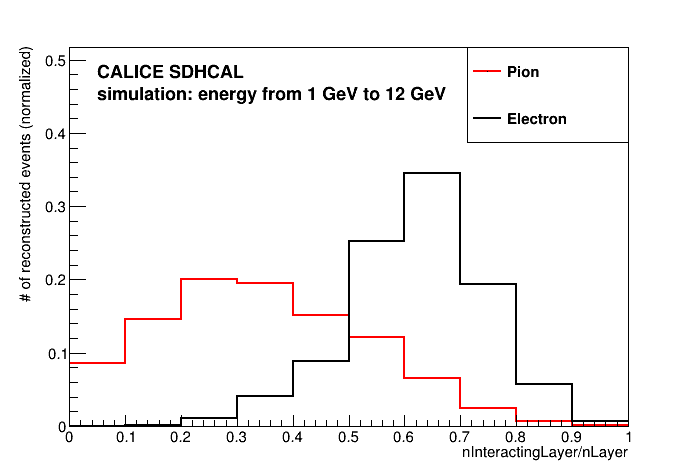}
\includegraphics[width=0.48\textwidth]{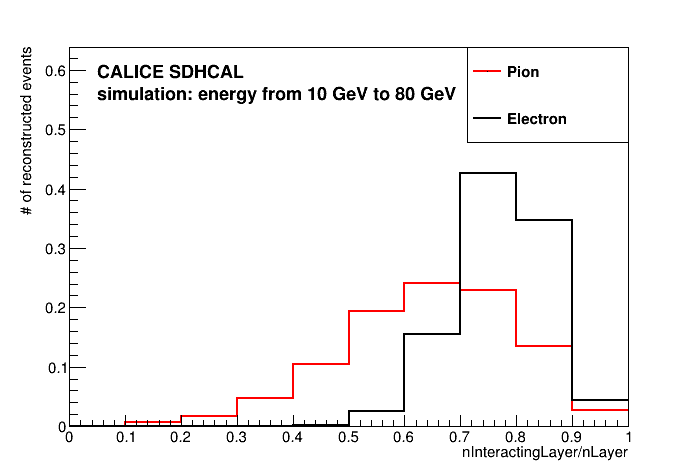}
\caption{\label{Ratio2} Distribution of the ratio of number of shower layers over the total number of fired layers in pion and electron showers in the PS energy range of 1-12~GeV (left) and the SPS energy range of 10-80~GeV (right) as given by the simulation. The red line corresponds to  pions and the  black one to electrons.}
\end{center}  
\end{figure}

\item {\bf The average number of hits per fired layers (nHit/nLayer)}: This is the ratio between the total number of fired pads over the number of layers with at least one fired pad. The distribution of this variable is shown in Fig.~\ref{Ratio1}. 

\begin{figure}[!h]
\begin{center}
\includegraphics[width=0.48\textwidth]{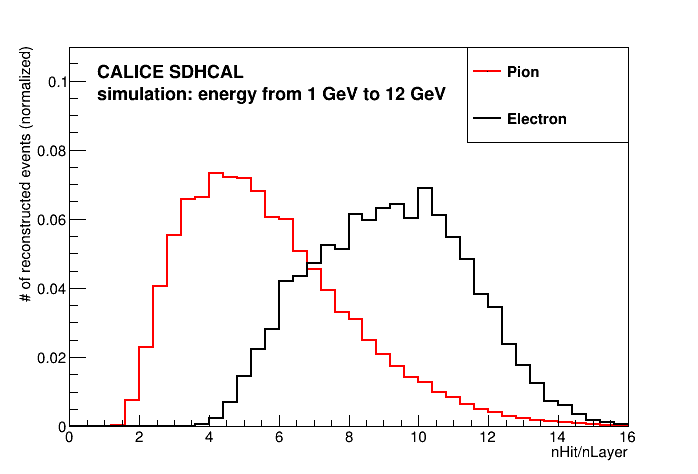}
\includegraphics[width=0.48\textwidth]{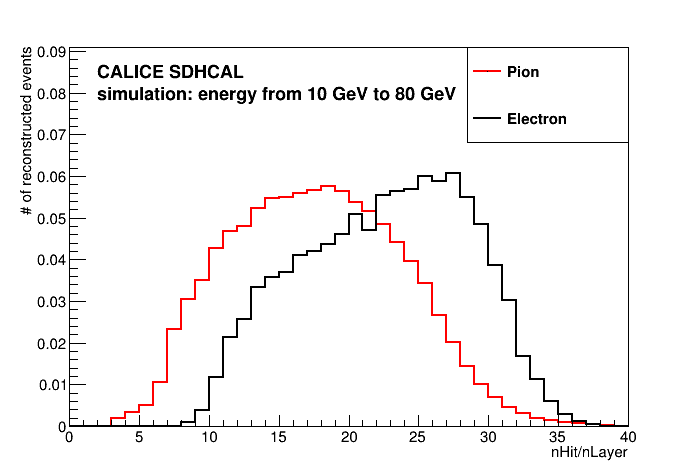}
\caption{\label{Ratio1} Distribution of the average number of hits per fired layers in pion and electron showers  in the PS energy range of 1-12~GeV (left) and  the SPS range of 10-80~GeV (right) as given by the simulation. The red line corresponds to pions and the black one to electrons.}
\end{center}  
\end{figure}

\item {\bf Shower density (Density):}  This is the average number of the neighbouring hits located  in the $3\times 3$ pads around one of the hits (including the hit itself) in the given event. Figure~\ref{density} shows clearly that electromagnetic showers are more compact than the hadronic ones as expected.

\begin{figure}[!h]
\begin{center}
\includegraphics[width=0.48\textwidth]{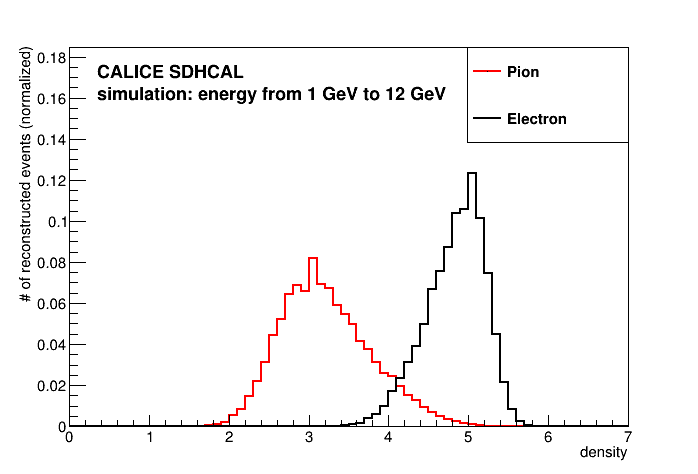}
\includegraphics[width=0.48\textwidth]{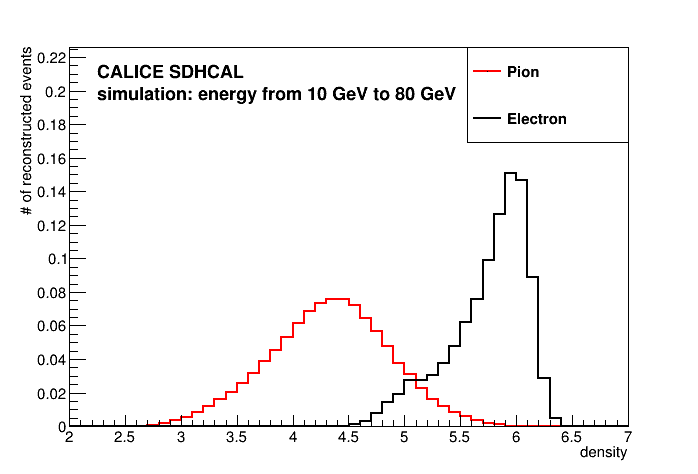}
\caption{\label{density} Distribution of the density of pion and electron showers  in the PS energy range of 1-12~GeV (left) and  the SPS range of 10-80~GeV (right)  as given by the simulation. The red line corresponds to pions and the black one to electrons.}
\end{center}  
\end{figure}

\item {\bf Shower radius (meanRadius):} This is the RMS of hits distance with respect to the event axis. To estimate the event axis, the average positions of the hits in each of the  ten first fired layers of an event are used to fit a straight line. The straight line is then used as the event axis. The electromagnetic shower being more compact than the hadronic shower, its radius is expected to be smaller as can be seen in Fig.~\ref{meanRadius}.

\begin{figure}[!h]
\begin{center}
\includegraphics[width=0.48\textwidth]{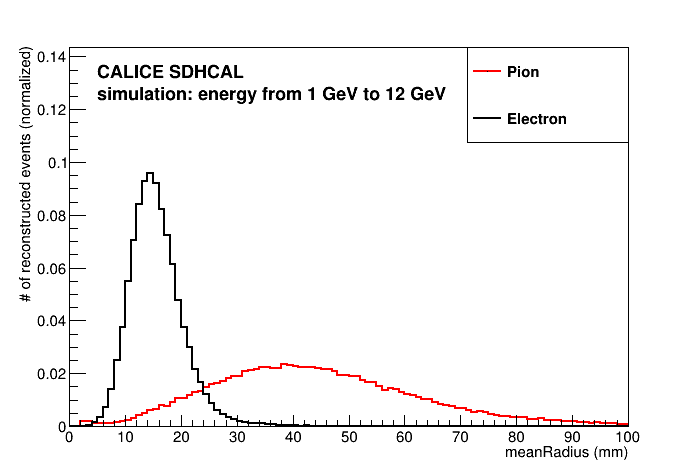}
\includegraphics[width=0.48\textwidth]{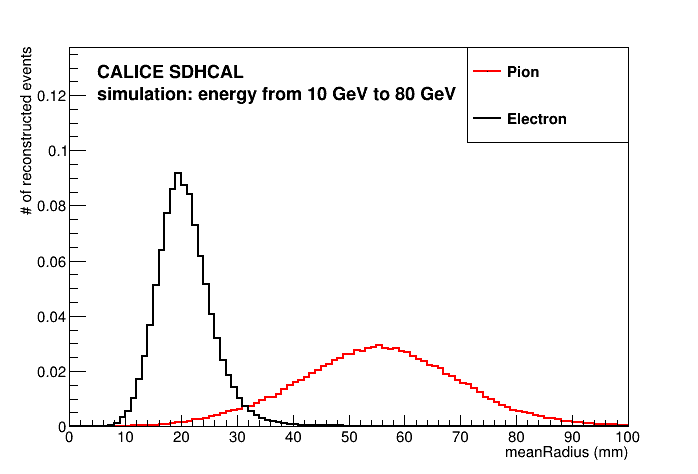}
\caption{\label{meanRadius} Distribution of the meanRadius of  pion and electron  showers in the PS energy range of 1-12~GeV (left) and  the SPS range of 10-80~GeV (right)  as given by the simulation. The red line corresponds to pions and the black one to electrons.}
\end{center}  
\end{figure}

\item {\bf Ratio of the number of third-threshold hits over the total number of hits (nHit3/nHit):} The three thresholds indicate the amount of charge collected in each pickup pad. The third one is set to single out the pads with high collected charge that may be induced by the passage of many particles in the cell associated to the pickup pad. The nHit3 is the number of third-threshold hits in one event. The ratio of nHit3 to the total number of hits helps to distinguish electromagnetic-like events and separate them from hadronic-like ones since the relative number of hits with the third threshold is higher in the former than in the latter due to the difference of their compactness. The distribution of this ratio can be seen in Fig.~\ref{nHit3}.
\end{itemize}
\begin{figure}[!h]
\begin{center}
\includegraphics[width=0.48\textwidth]{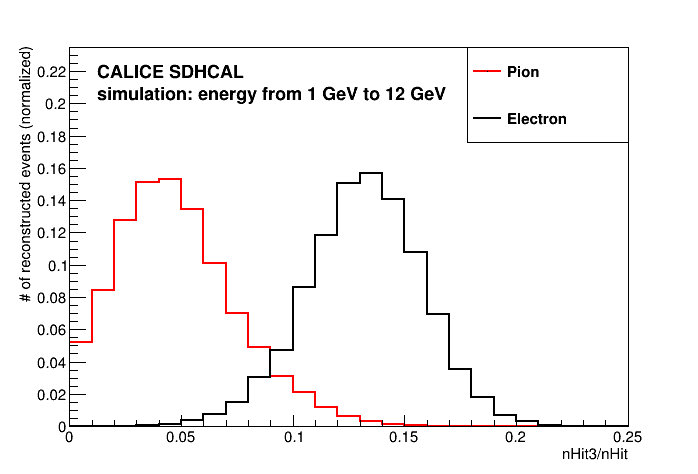}
\includegraphics[width=0.48\textwidth]{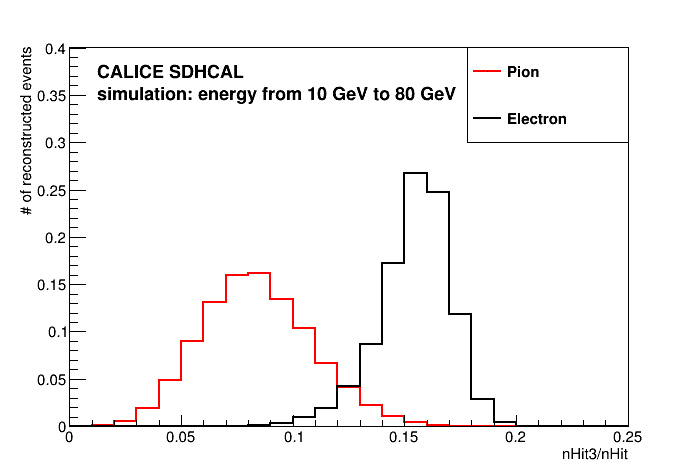}
\caption{\label{nHit3} Distribution of the ratio of number of third threshold hits over the total number of hits in pion and electron  showers in the PS energy range of 1-12~GeV (left) and  the SPS range of 10-80~GeV (right)  as given by the simulation. The red line corresponds to pions and the black one to electrons.}
\end{center}  
\end{figure}
\subsubsection{Training and testing details of the BDT}
\begin{figure}[!h]
\begin{center}
\includegraphics[width=0.65\textwidth]{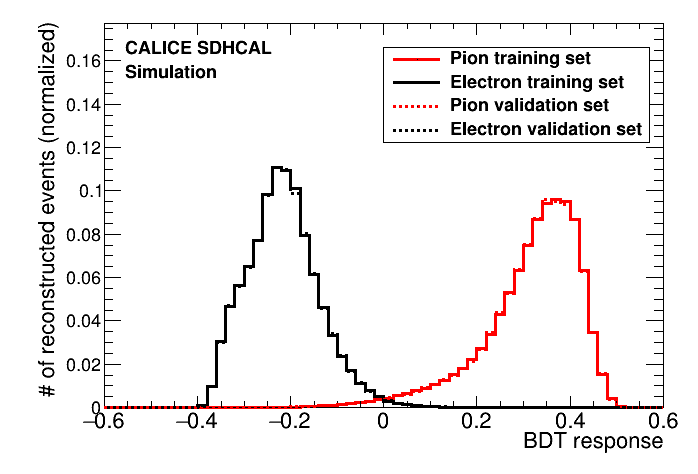}
\caption{\label{trainAndTest} Distribution of the BDT output of training and validation set using the simulated electron (black) and pion (red) events from 1 GeV to 80 GeV. The solid line is from training set while the dashed one is from validation set.}
\end{center}  
\end{figure}

\begin{figure}[!h]
\begin{center}
\includegraphics[width=0.45\textwidth]{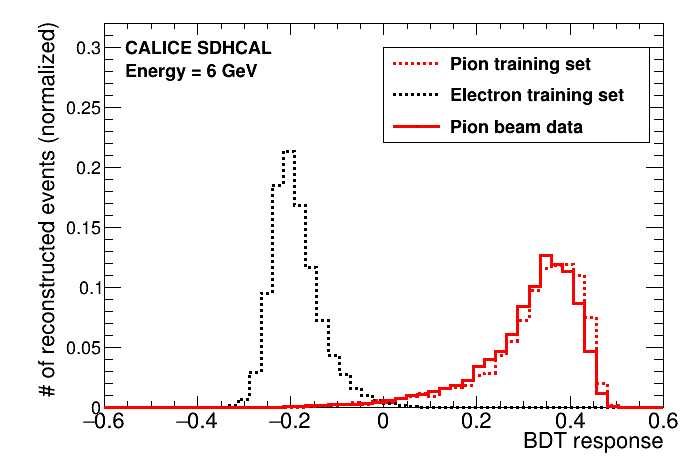}
\includegraphics[width=0.45\textwidth]{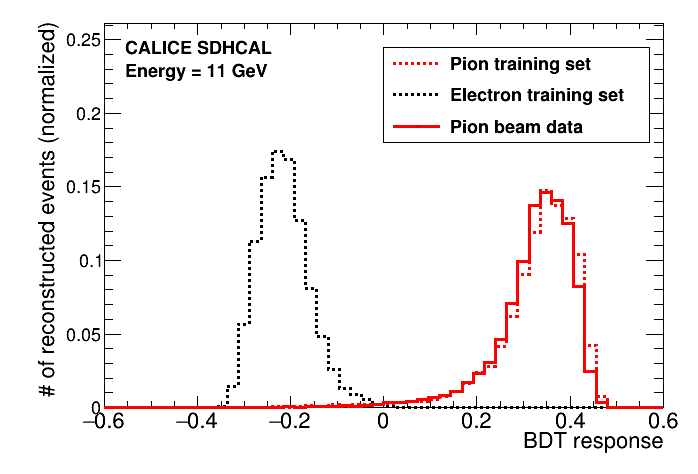}
\caption{\label{BDT_pie_6_11GeV} The BDT output of 6~GeV (left) and 11~GeV (right)  beam runs after muon rejection. The solid line is from pion beams and dashed one is from training set. }
\end{center}
\end{figure}

\begin{figure}[!h]
\begin{center}
\includegraphics[width=0.45\textwidth]{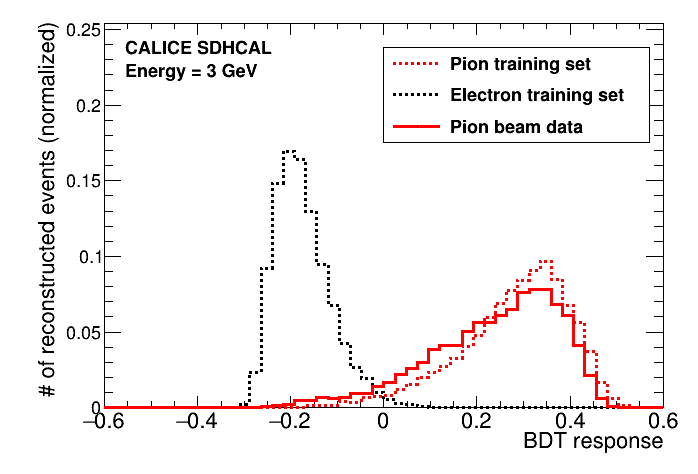}
\includegraphics[width=0.45\textwidth]{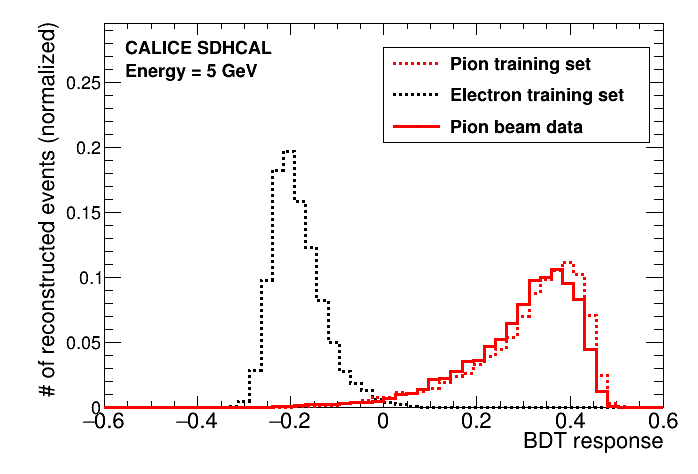}
\caption{\label{BDT_pie_3_5GeV} The BDT output of 3~GeV (left) and 5~GeV (right)  beam runs after muon rejection. The solid line is from pion beams and dashed one is from training set. }
\end{center}
\end{figure}

For the training and testing process, 200000 pion and 200000 electron  simulated events are used  to form a training set (66.7\%) and a testing one (33.3\%). Another independent 400000 events including pions and electrons are used as a validation set.  The events are simulated evenly in the energy interval between 1 and 80 GeV.
The hyperparameters resulting from the BDT optimisation procedure such as maxDepth\footnote{It controls the maximum depth of the tree that will be created. It can also be described as the length of the longest path from the tree root to a leaf. The root node is considered to have a depth of 0.} are described in Table~\ref{Tab:BDTPara}. 
\begin{table}
        \centering
        \caption{The chosen BDT hyperparameters.}
        \label{Tab:BDTPara}
        \begin{tabular}{|c|c|}
                \hline
                Option& Setting \\
                \hline
                Ntrees (Number of trees in the forest) & 1000 \\
                \hline
                nCuts (Number of steps during node cut optimisation) & 20\\
                \hline
                MaxDepth (Max depth of the decision tree allowed) & 4 \\
                \hline
        \end{tabular}
        
\end{table}
After feeding the eight topological variables to the BDT model using the training and testing sets, the performance of our model is shown in Fig.~\ref{trainAndTest}. It  clearly shows the strong separation power between pions and electrons. At the same time, the BDT response of the validation sets has very good agreement with training ones. This confirms that our model is performing very well and is not subject to overfitting.  After applying the muon rejection cuts to be explained in section to our in section~\ref{muon_rejection} to our collected data,  we apply our BDT models on the selected events.  Figure ~\ref{BDT_pie_6_11GeV} shows the BDT output of 6~GeV (left plot) and 11~GeV (right plot) pion runs which are supposed to be free of electron contamination.  The performance of the pion event selection  matches the one obtained with the simulated pion events quite well as shown in this figure, thus confirming that our model is reliable.  Figure~\ref{BDT_pie_3_5GeV} shows the result of 3~GeV and 5~GeV pion beam runs which, in principle, may contain electron contamination. From this figure, one can see that most of the events are located in the region associated to pions and a good agreement is observed between the data and the simulation even in the region of overlap between the pions and the electrons.  Therefore, the electron contamination in pion runs in the 3-5~GeV energy range is rather small even if it is more important than that of the PS higher energy runs. By requiring the BDT response to be larger than 0.0, we select pure pion events. 

\subsection{Muon contamination rejection}
\label{muon_rejection}
\begin{figure}[!h]
\begin{center}
\includegraphics[width=0.65\textwidth]{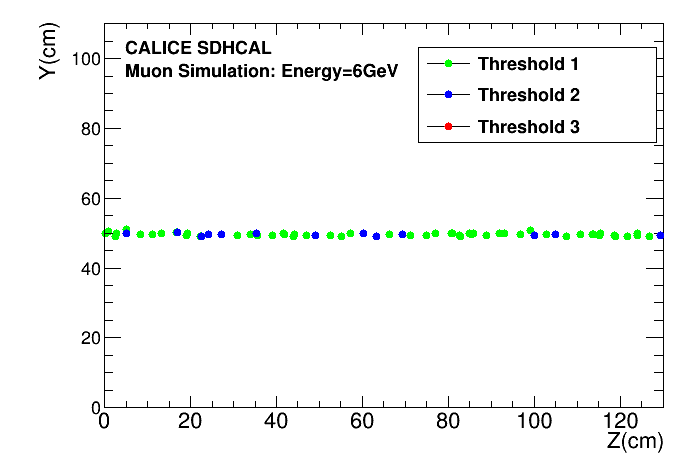}
	
\caption{\label{Muon} Event display of one 6~GeV simulated muon with the green, blue and red colour indicating the first, second and third threshold hits respectively. The third threshold is often absent in muon tracks because of the small amount of charge produced by muons in the RPC.}
\end{center}
\end{figure}

\begin{figure}[!h]
\begin{center}
\includegraphics[width=0.65\textwidth]{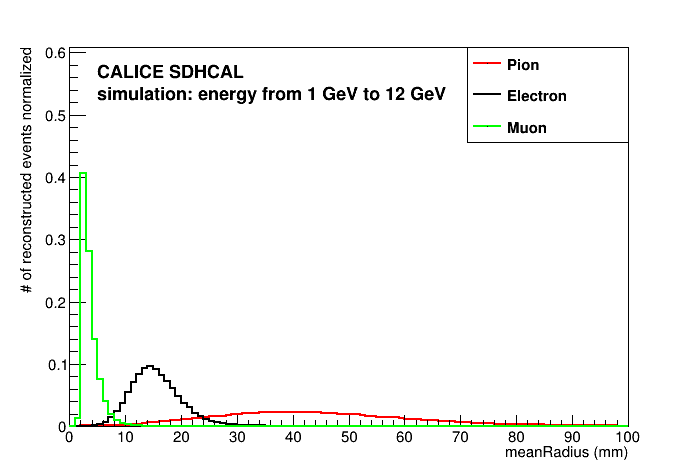}
\caption{\label{meanRadius_muon} Distribution of the meanRadius of shower by 1~GeV to 12~GeV muons as given by the simulation (green). Electrons(black) and pions(red) meanRadius distributions are also shown. }
 \end{center}
\end{figure}

The main contamination in our beam data is that of muons,  including beam muons and cosmic ones. To eliminate these two kinds of muons, we use the information based on the different behaviours  of muons and pions in the SDHCAL prototype. Basically,  muons  cross the prototype and only leave a straight track in the prototype like the one shown in Fig.~\ref{Muon}. The mean of hits distance (described by the variable meanRadius hereafter) of muon hits with respect to the global event axis is thus very often less than 1.5~cm ($\approx$ 1.5 pads) as shown in Fig.~\ref{meanRadius_muon}. 

%For the case of pions, when they come into the prototype, they produce hadronic showers with dense centers and sparse track segments like Figure ~\cite{Pion}.  So they will have more larger meanRadius than muons. 

To eliminate most of the muon contamination, we  require that the meanRadius is greater than 2~cm. To further reduce the muon contamination, including the so-called radiative muons that produce a few  hit clusters around the muon track, we require the ratio of the number of shower layers to the total number of layers with at least one hit  to be more than half.

\begin{figure}[!h]
\begin{center}
\includegraphics[width=0.65\textwidth]{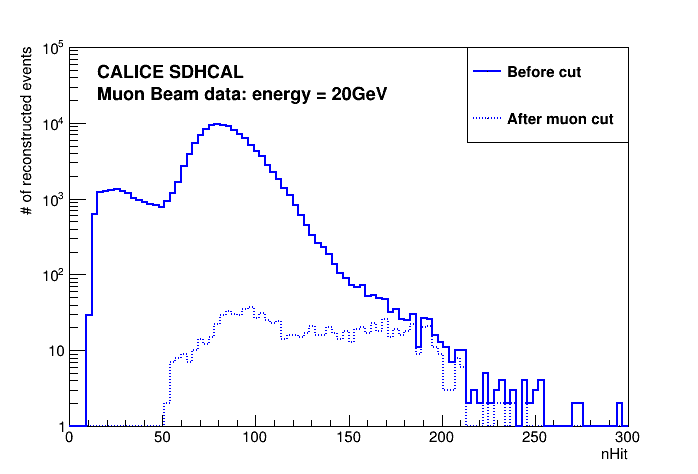}
\caption{\label{MuonCutData} Distribution of the number of hits for 20~GeV muon run before~(solid line) and after~(dashed line) muon cut.}
\end{center}
\end{figure}

 To check the rejection power of the muon cuts, we  apply it to dedicated muon runs.  Figure~\ref{MuonCutData} shows the distribution of the  number of hits before and after muon rejection for 120~GeV muon runs. It clearly shows the rejection power of this selection which is higher than 99.0\%. 

\begin{figure}[!h]
\begin{center}
\includegraphics[width=0.32\textwidth]{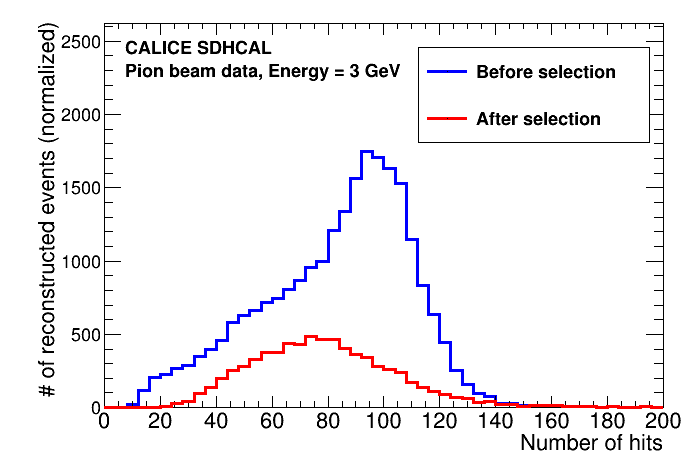}
\includegraphics[width=0.32\textwidth]{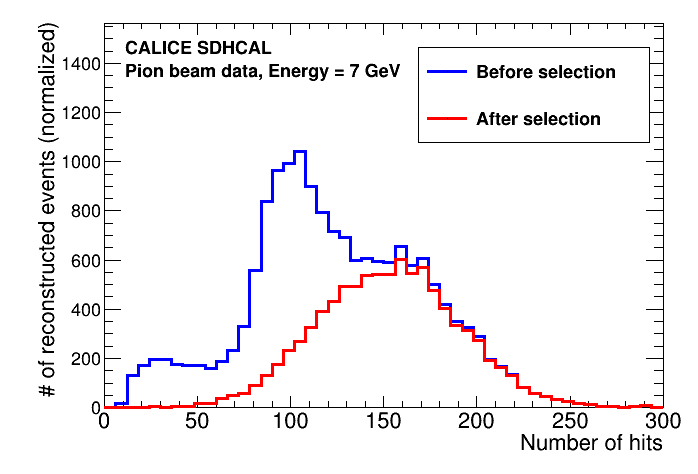}
\includegraphics[width=0.32\textwidth]{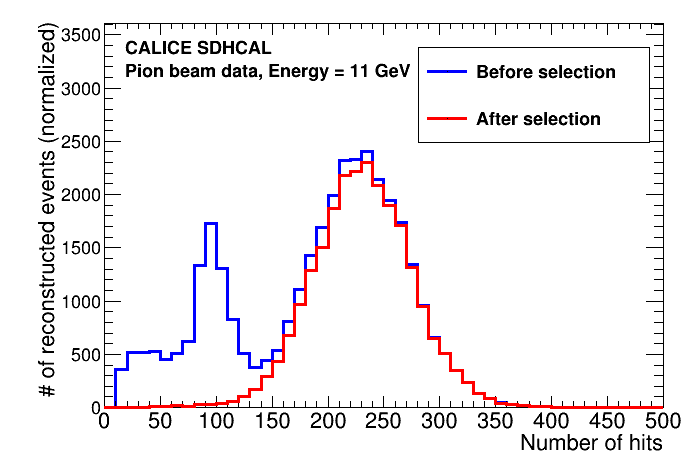}
\caption{\label{nHitResultPS} The number of hits for 3, 7 and 11~GeV pion beam runs before~(blue) and after~(red) muon selection. }
\end{center}
\end{figure}

\begin{figure}[!h]
\begin{center}
\includegraphics[width=0.32\textwidth]{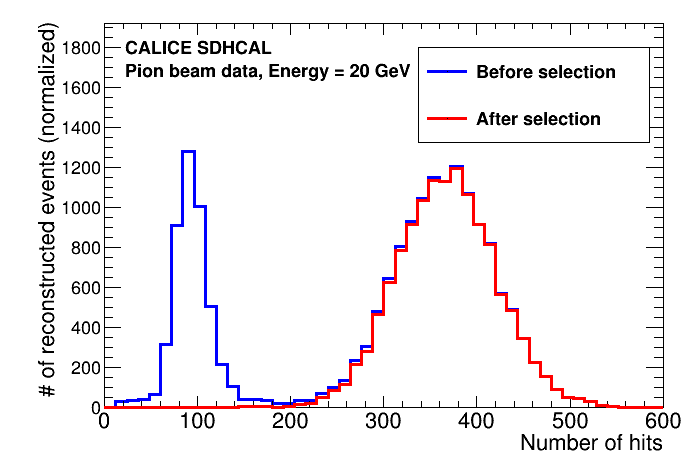}
\includegraphics[width=0.32\textwidth]{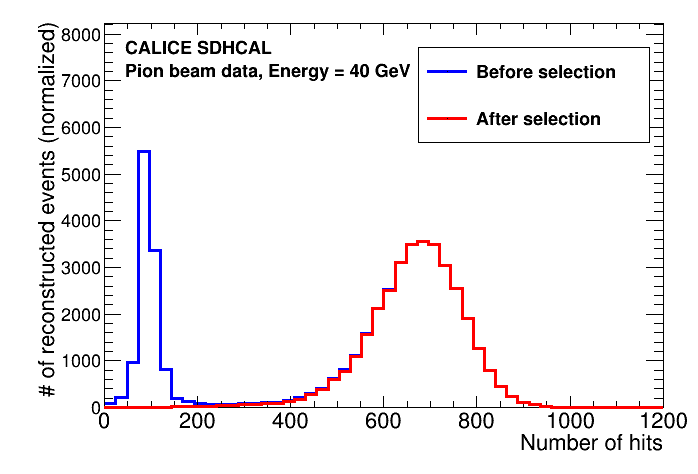}
\includegraphics[width=0.32\textwidth]{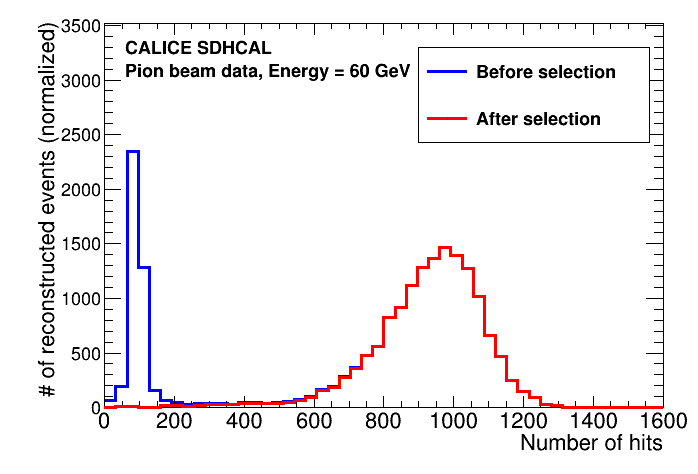}
\caption{\label{nHitResultSPS} The number of hits for 20, 40 and 60~GeV pion beam runs before~(blue) and after~(red) muon selection. }
\end{center}
\end{figure}

The result of the selection including the muon rejection and electron cut (BDT response > 0.0) is shown in Fig.~\ref{nHitResultPS} and  Fig.~\ref{nHitResultSPS} for PS and SPS beam data runs respectively. 

\section{Energy reconstruction}
\label{reco}

 % In addition to confirm that no bias is introduced, this figure shows that the SDHCAL response is stable under different beamline conditions. 

The rejection of electrons present in the pion data sample using the BDT but also that of the muons allows us to have pure pion sample as explained in the previous section.
The selected pion events of the PS beam energy from 3~GeV to 11~GeV and those in the range of 10-80~GeV of the SPS, can then be used to reconstruct energy.

%, combined with the high energy data samples (from 20~GeV to 80~GeV) taken at SPS in 2015~\cite{RefCAN}.
    Based on the information of the number of hits belonging to first threshold (nHit1), second threshold (nHit2) and third threshold (nHit3), the hadronic shower energy can be reconstructed as described in Ref.~\cite{FirstResults} using the following formula:
 \begin{equation}
E_{reco} = \alpha \times nHit1 + \beta \times nHit2 + \gamma \times nHit3
\label{eq:CalE}
\end{equation}
where $\alpha$, $\beta$ and $\gamma$ are weight factors which are parametrised as second order polynomials of the total number of hits $nHit= nHit1 + nHit2 + nHit3$:
\begin{equation}
\begin{split}
\alpha = \alpha_1 + \alpha_2 \times nHit + \alpha_3 \times nHit^2 \\
\beta = \beta_1 + \beta_2 \times nHit + \beta_3 \times nHit^2 \\
\gamma = \gamma_1 + \gamma_2 \times nHit + \gamma_3 \times nHit^2
\end{split}
\label{eq:Calpara}
\end{equation}
The nine parameters $\alpha_{i=1,2,3}$,  $\beta_{j=1,2,3}$ and $\gamma_{k=1,2,3}$ are obtained, as described in Ref.~\cite{FirstResults}, from  a  part of the data samples of a few energy points  by minimising the following $\chi^2$ expression:
\begin{equation}
\chi^2 = \sum_{i=1}^N \frac{(E_{beam}^i-E^i_{reco})^2}{\sigma^2_i}
\end{equation}
where the $E_{beam}^i$ denotes the beam energy and the $E^i_{reco}$ is the reconstructed energy. $N$ is the number of total events and  $\sigma_i = \sqrt{E_{beam}^i}$ where the choice of $\sigma =\sqrt{E_{beam}^i}$ is motivated by the fact that the expected energy resolution is approximately given by the stochastic term: $\frac{\sigma}{E_{beam}} = \frac{\alpha}{\sqrt{E_{beam}}}$.

\begin{figure}[!h]
\begin{center}
\includegraphics[width=0.65\textwidth]{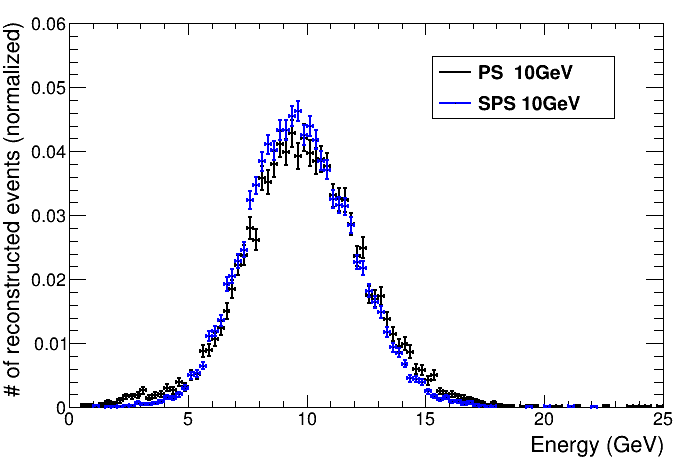}
\caption{\label{Ene-10} Reconstructed energy of the the PS 10~GeV pion sample  and that of the SPS 10~GeV sample after applying the BDT selection on the latter.}
\end{center}  
\end{figure}

Since the PS raw 10~GeV sample is almost free of electron contamination, it is therefore expected to be less impacted by the BDT-based selection. On the contrary,  the SPS raw 10~GeV sample electron  contamination is relatively higher and could thus be impacted by the BDT-based selection.   To check however that this selection only eliminates the electrons without changing the pion sample characteristics, the reconstructed energy of the  PS 10~GeV sample without BDT selection is compared to that of the SPS one after applying the BDT selection.  Fig.~\ref{Ene-10} shows the normalised reconstructed energy distribution of these two samples.  The  good agreement between the two distributions confirms the absence of bias of the  BDT selection and its efficiency in rejecting the electron contamination%~\footnote{The slight difference could be explained by the small difference of operation conditions between the two campaigns with respect to the pressure and temperature that were not corrected for, leading to slightly different gains and thus different numbers of hit.}.

\subsection{Energy resolution and linearity}
\label{resolin}

The two purified samples; the one of 3-11~GeV and the one of 10-80~GeV collected at the PS and the SPS beamline respectively, are then used to reconstruct the pion energy in the SDHCAL following the method described in Ref.~\cite{FirstResults}.

\begin{figure}[!h]
\begin{center}
\includegraphics[width=0.32\textwidth]{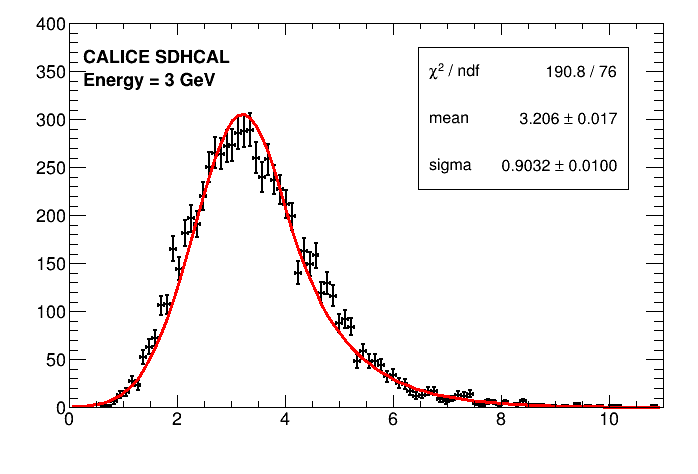}
\includegraphics[width=0.32\textwidth]{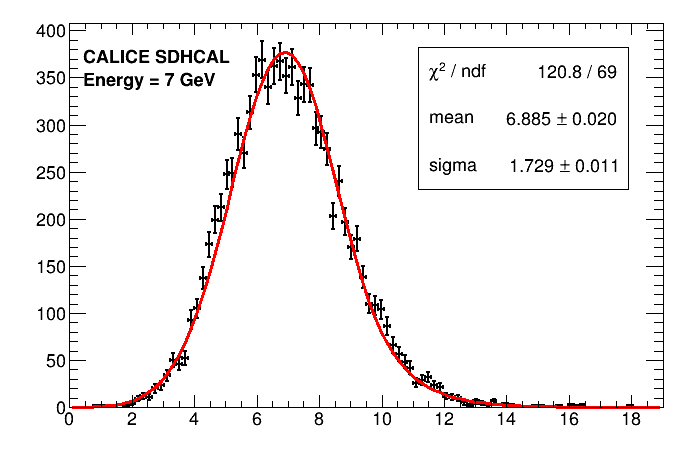}
\includegraphics[width=0.32\textwidth]{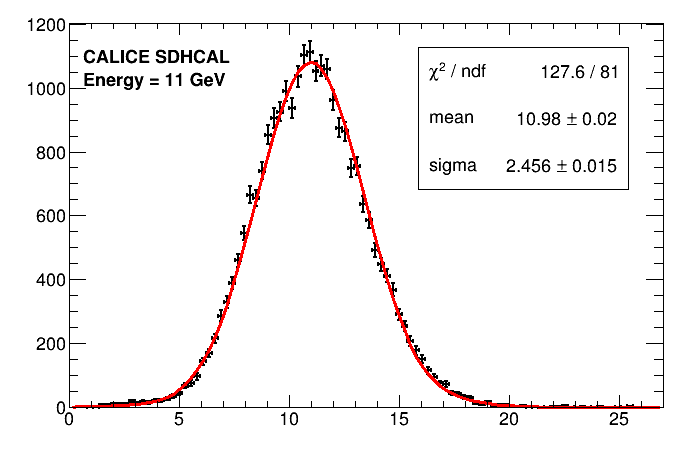}
\caption{\label{energyPS} Reconstructed energy distributions for 3~(left), 7~(middle) and 11~GeV~(right) pion data samples collected at the PS. The distributions are fitted with a double sided Crystal Ball function. The variance of the Gaussian part of the Crystal Ball function is used to estimate the resolution of the reconstructed energy.} 
\end{center}
\end{figure}

The reconstructed energy distributions of 3, 7 and 11~GeV pion data samples collected at PS are shown in Fig.~\ref{energyPS} .

\begin{figure}[!h]
\begin{center}
\includegraphics[width=0.32\textwidth]{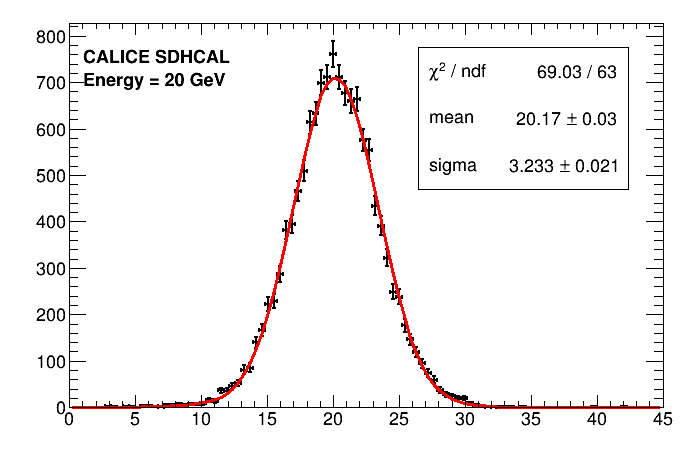}
\includegraphics[width=0.32\textwidth]{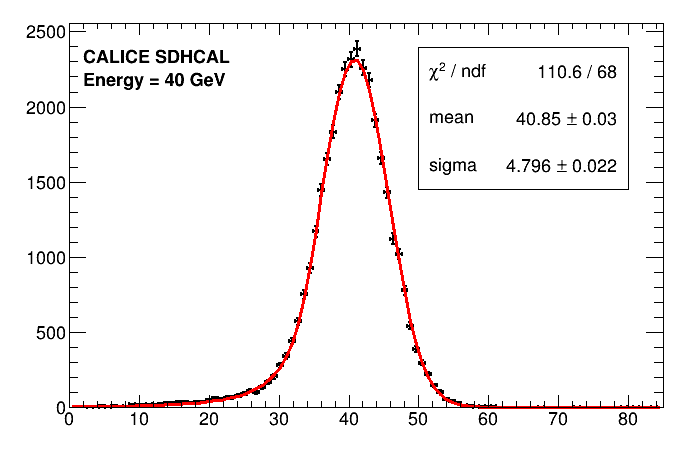}
\includegraphics[width=0.32\textwidth]{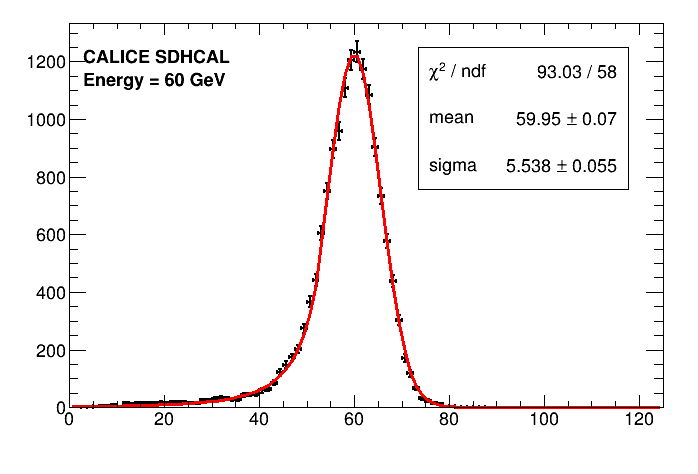}
\caption{\label{energySPS} Reconstructed energy distributions for 20~(left), 40~(middle) and 60~GeV~(right) pion data samples collected at SPS. The distributions are fitted with a double-sided Crystal Ball function. The variance of the Gaussian part of the Crystal Ball function is used to estimate the resolution of the reconstructed energy.}
\end{center}
\end{figure}

The reconstructed energy distributions of 20, 40 and 60~GeV pion data samples collected at SPS are shown in Fig.~\ref{energySPS} .

\begin{figure}[tbp]
\begin{center}
\begin{tabular}{cc}
\includegraphics[width=0.49\textwidth]{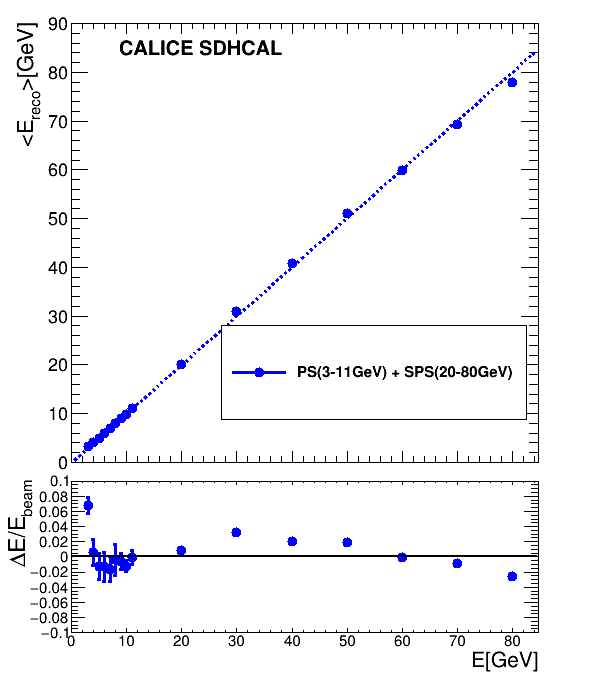}&
\includegraphics[width=0.49\textwidth]{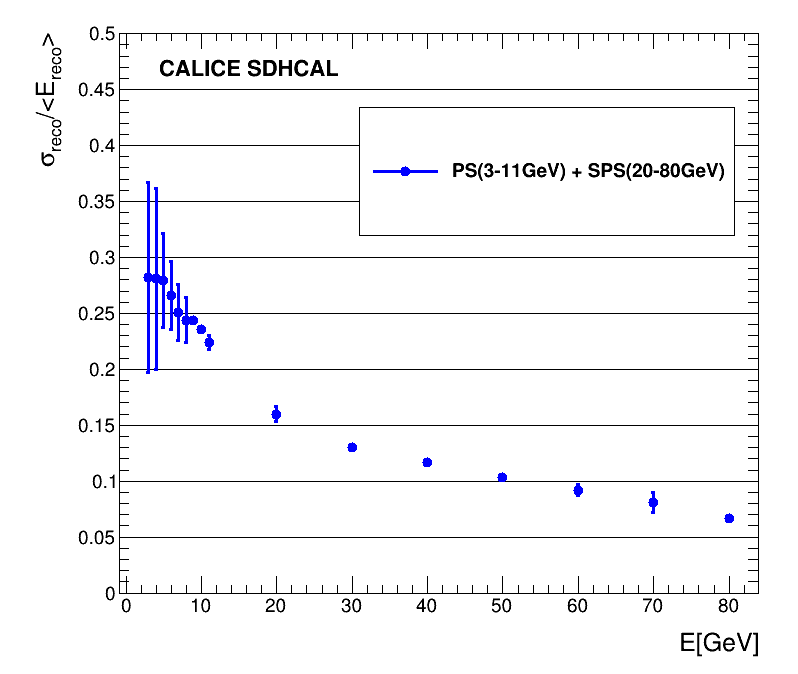}\\
 
\end{tabular}
\caption{Mean reconstructed energy of pion showers as a function of the beam energy as well as relative deviation of the pion mean reconstructed energy with respect to the beam energy (left) and resolution of the reconstructed hadron energy as a function of the beam energy (right). Both statistical and systematic uncertainties are included in the error bars. Dashed line on the left plot indicates the ideal linearity response of the calorimeter. }
\label{fig:Linearity_and_reso1}
\end{center}
\end{figure}

         After fitting the reconstructed energy distribution from 3 to 80~GeV,  the mean value and standard deviation of the Gaussian function are taken as the reconstructed energy and its resolution respectively. In  Fig.~\ref{fig:Linearity_and_reso1}, the energy linearity (left) and resolution results (right) are shown  using both  PS and SPS data.

% The energy resolution curve of both previous figures are fitted by the formula~\cite{RefCal}:
%\begin{equation}
%\frac{\sigma}{E} =  \frac{a}{\sqrt{E}} \oplus b
%\end{equation}
%where the $\oplus$ denotes the quadratic sum. The first term on the right-hand side is called the 'stochastic term', the %second term is the 'constant term'.  

\begin{figure}[tbp]
\begin{center}
\begin{tabular}{cc}
\includegraphics[width=0.49\textwidth]{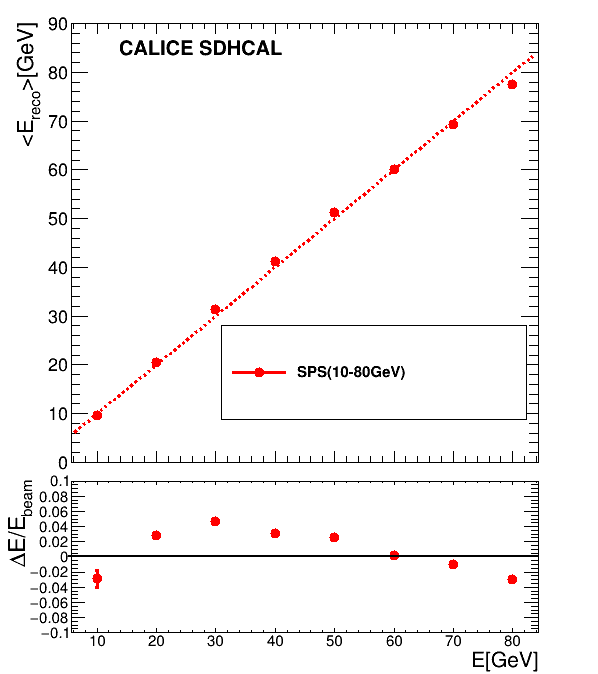}&
\includegraphics[width=0.49\textwidth]{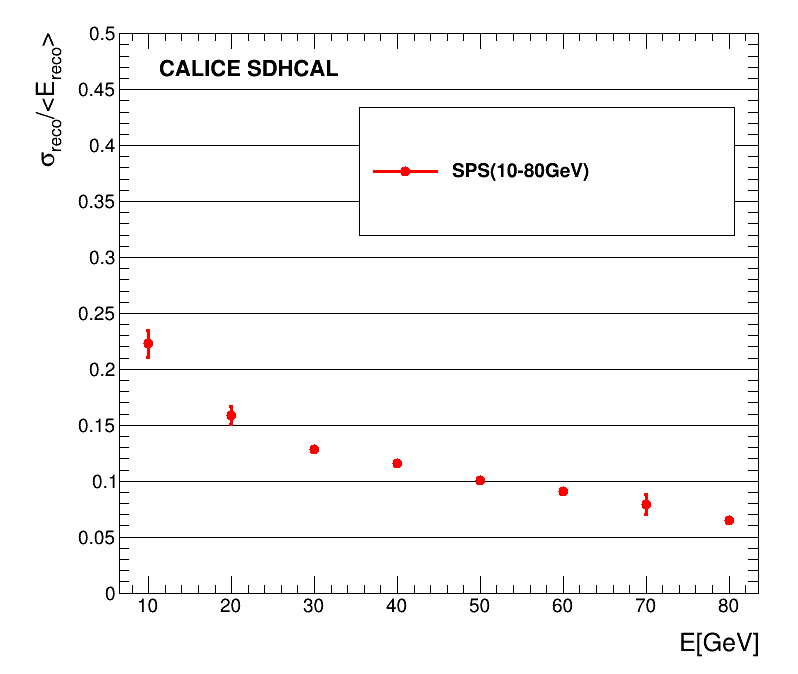}\\
 
\end{tabular}
\caption{Mean reconstructed energy for pion showers as a function of the beam energy as well as relative deviation of the pion mean reconstructed energy with respect to the beam energy (left) and resolution of the reconstructed hadron energy as a function of the beam energy (right). Both statistical and systematic uncertainties are included in the error bars. Dashed line on the left plot indicates the ideal linearity response of the calorimeter. }
\label{fig:Linearity_and_reso}
\end{center}
\end{figure}

The same procedure is applied to the SPS sample only.  Similar results as the one obtained with both PS and SPS beamlines are obtained as can be shown in Fig.~\ref{fig:Linearity_and_reso}.  More importantly, these  SPS results  are similar to those  obtained in 2012~\cite{FirstResults}.   This confirms the robustness of the SDHCAL prototype over time.

\section{Uncertainties estimation}
\label{uncertain}
The linearity and energy resolution results presented in section.~\ref{resolin} include statistical and systematic uncertainties. We present here after the main contributions to the systematic uncertainties:

\begin{itemize}

\item For the reconstructed energy of all energy points, a double sided Crystal Ball fit function and a Gaussian fit function are used. The difference of fitting results obtained from these two fit functions are considered as the value of systematic uncertainties associated to the fit of the reconstructed energy.
  %\item The difference of the estimated energy  before and after  applying the selection criteria is evaluated using simulation samples of pions from 3 to 80~GeV. The difference is used as one source of the systematic uncertainties.
  \item For the muon rejection, using all energy points data samples of PS, the meanRadius varied by an arbitrary 5\% in both directions with respect to the nominal values. The maximum deviation with respect to the nominal value is used as an estimate of the systematic uncertainties due to a residual muon contamination.
	\item For the electron rejection using the BDT method,  the BDT cut value is changed from -0.05 to 0.05 with respect to the nominal values 0.0. The maximum deviations are taken and added to the systematic uncertainties as an estimate of the impact of residual electron contamination.

\end{itemize}

Although the statistical uncertainties are found to be negligible for almost all the runs with respect to systematic uncertainties, their contributions as well as the systematic uncertainties previously discussed are added quadratically to obtain the final uncertainties. The results are summarised in Tables~\ref{tab:systematic_uncertainties_lin_low} and~\ref{tab:systematic_uncertainties_reso_low}. The uncertainty coming from the different fit functions is found to be  the main component of total systematic uncertainties. 

 \begin{table}[!ht]
\centering

\begin{tabular}{|c|c|c|c|}
\hline
Energy(GeV)  & Beam data of PS and SPS\\
\hline
$3 $ & $0.068\pm0.011 $ \\
\hline
$4$  & $0.006\pm0.017 $ \\
\hline
$5$  & $-0.013\pm0.017 $ \\
\hline
$6 $ & $-0.013\pm0.019 $  \\
\hline
$7$  & $-0.016\pm0.016 $\\
\hline
$8 $ & $-0.004\pm0.021 $  \\
\hline
$9$  & $-0.006\pm0.010 $ \\
\hline
$10  $&$ -0.012\pm0.007 $ \\
\hline
$11  $&$ -0.001\pm0.009 $ \\
\hline
$20 $ & $0.009\pm0.001 $ \\
\hline
$30$  & $0.032\pm0.001 $ \\
\hline
$40$  & $0.021\pm0.001 $ \\
\hline
$50 $ & $0.019\pm0.001 $  \\
\hline
$60$  & $-0.001\pm0.001 $\\
\hline
$70 $ & $-0.010\pm0.001 $  \\
\hline
$80$  & $-0.027\pm0.001 $ \\

\hline
\end{tabular}
\caption{\label{tab:systematic_uncertainties_lin_low}List of $\frac{\Delta E}{E}$ observed and associated uncertainties for beam data in the energy range from 3 to 80 GeV.}
\end{table}

\begin{table}[!ht]
\centering

\begin{tabular}{|c|c|c|c|}
\hline
Energy(GeV)  & Beam data of PS and SPS\\
\hline
$3$  & $0.282\pm0.085 $  \\
\hline
$4$  & $0.281\pm0.081 $  \\
\hline
$5$  & $0.279\pm0.042 $\\
\hline
$6 $ & $0.266\pm0.030$ \\
\hline
$7 $ & $0.251\pm0.025$  \\
\hline
$8 $ & $0.244\pm0.020$  \\
\hline
$9$  & $0.244\pm0.001$   \\
\hline
$10$  & $0.236\pm0.001$   \\
\hline
$11$  & $0.224\pm0.006$   \\
\hline
$20 $ & $0.160\pm0.007 $ \\
\hline
$30$  & $0.130\pm0.001 $ \\
\hline
$40$  & $0.117\pm0.001 $ \\
\hline
$50 $ & $0.103\pm0.002 $  \\
\hline
$60$  & $0.092\pm0.005 $\\
\hline
$70 $ & $0.081\pm0.009 $  \\
\hline
$80$  & $0.067\pm0.001 $ \\

\hline
\end{tabular}
\caption{\label{tab:systematic_uncertainties_reso_low}List of energy resolution observed and associated uncertainties for beam data in the energy range from 3 to 80 GeV.}
\end{table}

\section{Conclusion}
\label{con}

The data collected from the exposure of the SDHCAL prototype to pion beams  in both PS and SPS covering a large range of energy (3-80~GeV) are analyzed. Rejection of muon and electron contamination is performed. For the latter a BDT-based technique is applied.   This technique allows the rejection of the electron contamination without reducing the pion sample compared to the analysis used in Ref.~\cite{FirstResults} where the electron contamination was reduced by requiring the interaction to start showering after ten radiation lengths ($X_0$) leading to a loss of about half of the pion events. Energy of the pions collected in both PS and SPS is then reconstructed following the techniques developed in Ref.~\cite{FirstResults} and compared with those obtained with the SPS data only and with those obtained with those of 2012 following a standard selection analysis.  The results show that good performances including excellent linearity and energy resolution,  are obtained over a large dynamic range from 3 to 80~GeV.

\section{Acknowledgements}
This study was supported by the French ANR agency (DHCAL Grant) and the CNRS-IN2P3; by  the MCIN/AEI and the Programa Estatal de Fomento de la Investigaci\'on  Cientifica y T\'ecnica de Excelencia  Maria de Maeztu, grant MMDM-2015-0509;  by the National Natural Science Foundation of China (Grant No. 11961141006) and the 
  National Key Programmes for S\&T Research and Development (Grant No. 2016YFA0400404); by the National Research Foundation of Korea, Grant Agreement 2019R1I1A3A01056616.   The authors would also like to thank the CERN accelerator staff for their precious help in preparing both the PS and the SPS beams.

\end{document}